 \documentclass[twocolumn]{aastex6}

\usepackage{amssymb}	

\begin{document}

\title[Origin of peaked noise]
 {Interference as an origin of the peaked noise in accreting X-ray binaries}

\author{Alexandra~Veledina\altaffilmark{1,2,3}}
\affil{$^1$Nordita, KTH Royal Institute of Technology and Stockholm University, 
                             Roslagstullsbacken 23, SE-10691 Stockholm, Sweden\\
        $^2$Tuorla Observatory, University of Turku, V\"ais\"al\"antie 20, FI-21500 Piikki\"o, Finland; alexandra.veledina@gmail.com\\
        $^3$Nordita Fellow}

\begin{abstract}
We propose a physical model for the peaked noise in the X-ray power density spectra of accreting X-ray binaries.
We interpret its appearance as an interference of two Comptonization continua: one coming from the up-scattering 
of seed photons from the cold thin disk and the other fed by the synchrotron emission of the hot flow.
Variations of both X-ray components are caused by fluctuations in mass accretion rate, but there is a delay between them corresponding to the propagation timescale 
from the disk Comptonization radius to the region of synchrotron Comptonization.
If the disk and synchrotron Comptonization are correlated, the humps in the power spectra are harmonically related and the dips between them appear at 
frequencies related as odd numbers 1:3:5.
If they are anti-correlated, the humps are related as 1:3:5, but the dips are harmonically related.
Similar structures are expected to be observed in accreting neutron star binaries and supermassive black holes.
The delay can be easily recovered from the frequency of peaked noise and further used to constrain the combination of the viscosity parameter and 
disk height-to-radius ratio $\alpha (H/R)^2$ of the accretion flow.
We model multi-peak power spectra of black hole X-ray binaries GX~339--4 and XTE~J1748--288 to constrain these parameters.
\end{abstract}


\section{Introduction}

Astrophysical black holes can be studied by both spectral and timing techniques.
Spectral approach was more used in the beginning of the X-ray era due to the typically little amount of X-ray photons.
Temporal properties on short timescales, down to milliseconds, are perhaps the most challenging to probe.
A revolution in the field of black hole astrophysics was made by the X-ray mission {\it Rossi X-ray Timing Explorer}, with the help of which a vast amount 
of information on Galactic compact objects was collected.
It was realized that many properties of these objects can be understood by studying their variability patterns, which depend greatly on the spectral state and black hole mass.

The X-ray power spectral densities (PSDs) of black hole binaries generally demonstrate aperiodic variability and the low-frequency 
quasi-periodic oscillations \citep[QPOs,][]{CBS05}.
Early studies suggested that the hard-state broadband PSD constitutes a double-broken power-law  with a $f^{-1}$ dependence ($f$ is the Fourier frequency) 
in the range $\sim0.01-10$~Hz, below and above this range the variability is suppressed \citep{NGM81,BH90,CGR01}.
This $f^{-1}$ power-law was successfully reproduced in the model of propagating fluctuations of mass accretion rate \citep{Lyub97,KCG01,AU06}.
A model was extended by studies of fluctuations damping \citep{ZKM09} and modeling of the QPO \citep{IDF09,ID11,VPI13}.
It was also examined with global accretion disk simulations \citep{HR15}.
Yet all these investigations obtain an originally proposed power-law $f^{-1}$ in the frequency range of interest.
The PSDs obtained in propagating fluctuations model can now be directly fitted to the observed PSDs \citep{ID12,IvdK13}, but they tend to have somewhat steeper 
slope than the observed one in the frequency range $0.1-1$~Hz, as the data seems to be more consistent with two discrete peaks rather than the flat noise produced 
in this model \citep{BPvdK02,PWN03,ABL05,GPB14}.
Many studies suggest that the $f^{-1}$ power-law is a very crude approximation to the variability spectrum, as the PSD often shows a number of distinct peaks (the peaked noise components), which are usually fitted with Lorentzians \citep{NWD99,RTB00,Homan01,BPvdK02,Belloni06}.
A possible interpretation is that certain radii of the accretion flow are somewhat more variable than others.
For instance, the enhancement of variability is expected in the transitional region between the cold disc and hot flow \citep{VCG94b,CGR01}, or at the hot flow inner boundary 
if the orbital and black hole spins are misaligned \citep{F09,ID12}.
Admittedly, more than forty years after the first publication of X-ray PSD from Cyg~X-1 \citep{Terrell72}, no physical model was proposed 
to explain the complex multi-peak structure.

\begin{figure}
\figurenum{1}
\plotone{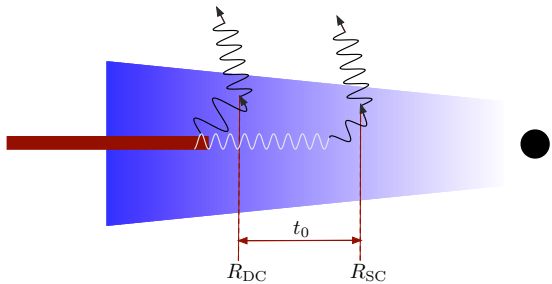}
\caption{
Sketch of the proposed geometry.
The horizontal oscillations (white) show the propagating fluctuations of mass accretion rate, which are eventually transformed to radiation (black lines) that is 
up-scattered to X-rays through disk Comptonization (at radius $R_{\rm DC}$) and synchrotron Comptonization (at radius $R_{\rm SC}$).
The light-curves produced by these two mechanisms should be similar, apart from damping of high frequencies in the disk Comptonization term and 
possible anti-correlation between them.
There is a time delay between them, corresponding to the characteristic time required for the fluctuations to propagate from $R_{\rm DC}$ to $R_{\rm SC}$.
}\label{fig:geometry} 
\end{figure}

In this work we propose the first physical explanation for the peaked noise in the PSD of black hole binaries.
We suggest that it is produced by the interference of two X-ray components, one being the  disk Comptonization and the other being synchrotron Comptonization.
Both components are variable as a result of propagating mass accretion rate fluctuations, both having the same PSD except for the possible damping at some frequency.
An essential requirement is that there is a time delay between them, corresponding to the propagation time between the disk Comptonization radius, 
likely close to the cold disk truncation radius, and the radius where the synchrotron is produced and Comptonized within the hot accretion flow.

Such a scenario is supported by the spectral data. 
The cold disk is commonly believed to be the source of seed photons in the soft state \citep{DGK07}, but a number of arguments suggest alternative sources of seed photons 
in the hard state, such as synchrotron emission self-generated in the flow \citep{Esin97,PV09,MB09,VPV13,PV14,YN14}.
It is then likely that we see contributions of both sources in the intermediate state.

In this paper we discuss dependence of resulting PSD shape on the model parameters and show that many observable features can be understood in terms of this mechanism.
We apply the developed model to the multi-peak PSDs observed in black hole binaries GX~339--4 and XTE~J1848--288.
We show that the delay time can be easily recovered from the PSD shape, with the help of which we can put constraints 
on the combination of disk height-to-radius ratio and viscosity parameter $\alpha (H/R)^2$, which is not measured directly for the hot flow.

\begin{figure*}
\figurenum{2}
\plotone{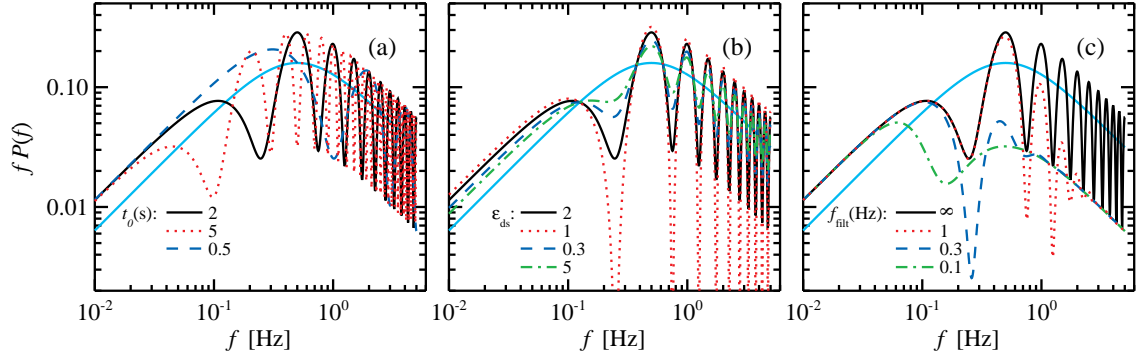}
\caption{
Power spectra obtained from the model.
Black solid line in every panel corresponds to the fiducial parameter set ($t_0=2$~s, $\varepsilon_{\rm ds}=2$, no damping).
Cyan solid line in every panel is the seed Lorentzian, plotted for comparison purposes.
Panel (a): dependence on the delay time $t_0$.
Panel (b): dependence on $\varepsilon_{\rm ds}$.
Panel (c): dependence on the damping frequency $f_{\rm filt}$.
Parameters for each curve can be found in Table~\ref{tab:par}.
}\label{fig:model_params} 
\end{figure*}

\section{Model}\label{sect:model}

We consider a scenario where the disk is truncated at some radius and the hot flow exists within this truncation radius \citep{PKR97,Esin97}.
In this geometry, the mass accretion rate fluctuations are excited at the cold accretion disc and are propagated from its inner boundary into the hot flow.
The excited fluctuations lead to variations in the disc seed photons on the dynamical timescales at the truncation radius.
The disc seed photons penetrate into the hot flow, where they are being Compton up-scattered to the X-rays.
In addition to this, fluctuations in mass accretion rate $\dot{m}$ propagate from the disk truncation radius into the hot flow, exciting the X-ray fluctuations again, 
this time via synchrotron Comptonization.
The sketch of the considered geometry is shown in Figure~\ref{fig:geometry}.

Disc photons are effectively Comptonized in the region close to the truncation radius, however, 
not many disc photons penetrate to the synchrotron Comptonization region due to a small dilution factor.
On the other hand, the dilution factor for synchrotron photons is not so small and they presumably may travel within the flow.
However, the optical depth of the flow is of the order of unity (as commonly suggested by the spectral fitting, \citealt{ZG04}), 
so the emitted synchrotron photons are also likely to be Comptonized locally.

We assume that the observed fluxes produced by two mechanisms (Comptonization of the disk and the synchrotron photons) are directly proportional to the 
mass accretion rate and are additive, so the total X-ray light-curve can be described by
\begin{equation}\label{eq:x_noqpo}
 x(t) = \left[\varepsilon_{\rm ds}\dot{m}(t+t_0)\ast h(t) + \dot{m}(t)\right]/\sqrt{1+\varepsilon_{\rm ds}^2},
\end{equation}
where $\dot{m}$ denotes the light-curve of mass accretion rate fluctuations (throughout the paper, we use small letters for quantities in time domain and 
capital letters for frequency domain).
The first term of this equation describes the emission coming from disk Comptonization,
while the second term corresponds to the synchrotron Comptonization.
The parameter $\varepsilon_{\rm ds}$ regulates relative importance of the two terms.
The denominator is the normalization factor.
There is a time delay $t_0$ between the time disk photons Comptonize and the time the synchrotron photons Comptonize within the flow, 
corresponding to a characteristic propagation time from the truncation radius to the region where synchrotron Comptonization operates.
This latter region might be extended along the radius, leading to some smearing of the variability from the synchrotron Comptonization light-curve
compared to the mass accretion rate light-curve.
However, the characteristic smearing time, of the order of viscous timescale at the synchrotron Comptonization radius, is much less that the delay 
time $t_0$, so the $\dot{m}(t)$ is still a good approximation for the synchrotron Comptonization light-curve.
The disk Comptonization component is variable at the dynamical timescale of the truncation radius.
It is lacking high-frequency signal, as it is originating at radii much farther than the characteristic radius of synchrotron Comptonization.
This is modeled by a convolution ($\ast$ sign) of $\dot{m}(t)$ with the filter function $h(t)$.
Its Fourier image is taken in the form
\begin{equation}\label{eq:filter_func}
 H(f) = \frac{1}{(f/f_{\rm filt})^4+1},
\end{equation}
where $f_{\rm filt}$ is the characteristic damping/filtering frequency.

The broadband X-ray PSD corresponding to the light-curve from Equation~(\ref{eq:x_noqpo}) is described by
\begin{equation}\label{eq:xpsd_cross}
 |X|_{\rm bb}^2(f) = \frac{ 1 + \varepsilon_{\rm ds}^2H^2(f) + 2\varepsilon_{\rm ds} H(f)\cos{(2\pi f t_0)} }{1+\varepsilon_{\rm ds}^2} |\dot{M}|^2(f),
\end{equation}
where the cross-term proportional to $\cos{(2\pi f t_0)}$ appears.
This produces oscillations in the X-ray PSD and, depending on the value $\varepsilon_{\rm ds}$, can significantly reduce power at frequencies $f=(k-1/2)/t_0$, 
where $k$ is an integer number.
The peaks appear at $f=k /t_0=kf_0$. 
Hence, $f_0$ can be considered as fundamental frequency and the other peaks as harmonics.
The characteristic width of the peak set by the cosine term is $\delta f=2[(k+1/4)/t_0-k/t_0]=1/(2t_0)$, hence the peak quality factor $f/\delta f$ 
increases with increasing frequency.
We note that the cross-term only redistributes the power between frequencies, the integral is preserved because the cosine dependence vanishes after integration.

\begin{table}
\caption{Parameters of numerical modelling for Figure~\ref{fig:model_params}.
}\label{tab:par}
  \begin{center}
\begin{tabular}{ccccc}
\hline
\hline
Panel 	& Line	& 	$t_0$~(s)	& $\varepsilon_{\rm ds}$	&$f_{\rm filt}$~(Hz) 	\\
\hline 
 a,b,c  	& black solid		&	2		&	2		&	$\infty$	 	\\
 a		& red dotted		&	5	 	&      2   		&	$\infty$	  \\
 a		& blue dashed		&	0.5	 	&      2    		&	$\infty$	  \\
 b		& blue dashed		&	2	 	&      0.3   		&	$\infty$	  \\
 b		& red dotted		&	2	 	&      1    		&	$\infty$	  \\
 b		& green dot-dashed	&	2	 	&      5    		&	$\infty$	  \\
 c		& green dot-dashed	&	2	 	&       2   		&	0.1	  \\
 c		& blue dashed		&	2	 	&      2    		&	0.3	  \\
 c  		& red dotted		&	2	 &       2   		&	1.0	  \\
\hline
     \end{tabular}
  \end{center}
\end{table}

\subsection{Dependence on parameters}

The study of the dependence of the PSD shape on model parameters can be done analytically using Equation~(\ref{eq:xpsd_cross}).
As an illustration, we use a single zero-centered Lorentzian function to describe the shape of the mass accretion rate PSD
\begin{equation}
 |\dot{M}|^2(f)={\cal L}(f)\equiv\frac{\Delta f/\pi }{(\Delta f)^2 + f^2}, 
\end{equation}
where we take the characteristic width $\Delta f=0.5$~Hz.
The resulting PSDs are shown in Figure~\ref{fig:model_params} and the parameters for each curve are listed in Table~\ref{tab:par}. 
The increase of time delay results in an increase of the number of humps, also the strength of the harmonic peaks relative to the low-frequency bump increases.	
A more pronounced cross-term (oscillations in PSD) appears when the parameter $\varepsilon_{\rm ds}$ approaches 1, the oscillations disappear both 
at  $\varepsilon_{\rm ds}\gg1$ and $\varepsilon_{\rm ds}\ll1$.
The behavior is easily understood from Equation~(\ref{eq:xpsd_cross}), suggesting that an interference between the two X-ray components is mostly seen when they have
similar amplitudes and the effect vanishes once one of them dominates.
This also implies that changing $\varepsilon_{\rm ds}$ to $1/\varepsilon_{\rm ds}$ does not alter the PSD shape (here we assume no damping, $H(f)=1$), 
as the importance of the cosine term is  determined by the ratio $2\cos{(2\pi f t_0)}/(\varepsilon_{\rm ds}+1/\varepsilon_{\rm ds})$.
The dependence on damping frequency is rather straightforward: the smaller $f_{\rm filt}$ is, the less the effects of the cross-term are and the less oscillations the PSD shows.
Interestingly, for low filtering frequency, the PSD shows the double-hump shape, which resembles those often observed \citep[e.g.,][]{BPvdK02}.

\begin{figure}
\figurenum{3}
\plotone{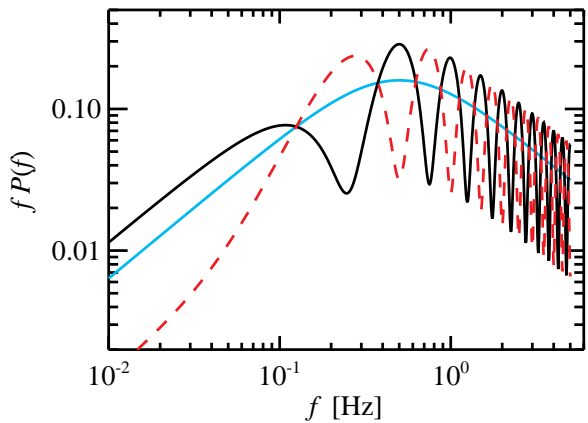}
\caption{
Power spectra obtained from the model for different signs of $\varepsilon_{\rm ds}$.
Parameters for the black solid line can be found in Table~\ref{tab:par}.
The parameter $\varepsilon_{\rm ds}=-2$ for the red dashed line, other are the same as for the black line.
Seed Lorentzian is shown with cyan solid line.
}\label{fig:anticorr} 
\end{figure}

\subsection{Different dependence on $\dot{m}$}

So far we assumed that variations of both disk and synchrotron Comptonization correlate with mass accretion rate fluctuations, however, this might not be realized 
for X-rays in some energy range.
For instance, the disk Comptonization can produce spectral pivoting, as the amount of photons at lower energies are correlated with $\dot{m}$, while the amount 
of photons at higher energies is anti-correlated due to a decrease of the electron temperature \citep[as reported, e.g., in][]{ZGM04}.
The variations of $\dot{m}$ may also cause spectral pivoting in the synchrotron Comptonization \citep{VPV11}, light-curves of photons with energies above 
pivoting point are correlated with $\dot{m}$, but for photons with lower energies they are anti-correlated with $\dot{m}$.
The behavior of these two components can be simply described as ``softer when brighter'' for the disc- and ``harder when brighter'' for synchrotron Comptonization.
Depending on the position of the pivoting point and the energy range considered, both components can be seen correlated with accretion rate, anti-correlated with 
accretion rate (in these two cases they are correlated with each other) or one is correlated, the other is anti-correlated with $\dot{m}$ 
(in which case they are anti-correlated with each other).

Interestingly, two distinct variability patterns, with constant and changing spectral slope, were reported for Cyg~X-1 \citep{ZPP02}.
Though found for the long-term variations, the behavior might be similar in the short-term variability, such as the ``harder when brighter'' pattern reported for 
GX~339--4 \citep{GMD08}, and the ``softer when brighter'' behavior found in XTE~J1118+480 \citep{MBSK03}, suggesting Comptonization of synchrotron 
and disk to be dominant, respectively.
 
We investigate how the PSD changes if two sources are anti-correlated.
This can be modeled by changing the sign of parameter $\varepsilon_{\rm ds}$.
The resulting PSDs are shown in Figure~\ref{fig:anticorr}.
As can be easily shown analytically, the dips and peaks in PSD are now reversed.
The peaks appear at frequencies $f=(k-1/2)/t_0$ and are related as odd numbers 1:3:5, while the dips are harmonically related and appear at $f=k/t_0$.
The PSD shape below $\sim0.05$~Hz is different, now the PSD function is convex (its second derivative is positive), while for cases considered earlier it 
had a concave shape (negative second derivative).

\begin{figure}
\figurenum{4}
\plotone{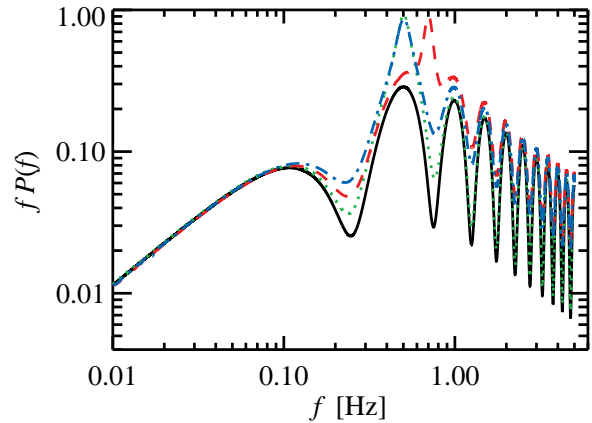}
\caption{
Model power spectra with additional contribution of the QPO.
Parameters for black solid line can be found in Table~\ref{tab:par}.
The QPO parameters are: $\Delta f_{\rm QPO}=0.05$~Hz, $\varepsilon_1=1$, $f_{\rm QPO}=0.7$~Hz (red dashed line) and $0.5$ (blue dot-dashed line), other 
parameters are the same as for the black line.
The PSD with the QPO coupled to the broadband variability in an additive way is shown with the green dotted line ($f_{\rm QPO}=0.5$~Hz), we note the difference 
in oscillation amplitude in this case.
The PSDs are renormalized to match at low frequencies.
}\label{fig:qpo} 
\end{figure}

\subsection{Coupling with QPO}\label{sect:qpo}

Broadband variability in almost every X-ray PSD is accompanied by the low-frequency QPO.
We assume that the QPO is coupled to the aperiodic variability in a multiplicative way, i.e. it modulates the accretion rate fluctuations. 
In this case, no simple analytical approximation can be obtained for the resulting PSD shape, so we proceed with the simulations of the X-ray light-curves. 
The quantity $\dot{m}(t)$ is obtained from its PSD (zero-centered Lorentzian with $\Delta f=0.5$~Hz as in previous cases) using the \citet{TK95} algorithm, with zero mean.
The QPO PSD has a form of a narrow Lorentzian
\begin{equation}\label{eq:qpo_lor}
 {\cal Q}_n(f)=\frac{\Delta f_{{\rm QPO},n}/\pi}{(\Delta f_{{\rm QPO},n})^2 + (f-nf_{\rm QPO})^2},
\end{equation}
where subscript $n$ corresponds to a harmonic number. 
In this section, we consider only one harmonic.
The QPO light-curves are obtained from this Lorentzian also using the \citet{TK95} algorithm, with zero mean.

We model the total light-curve as
\begin{equation}\label{eq:x_qpo}
 x(t) = \varepsilon_{\rm ds}\dot{m}_{\rm filt}(t+t_0) + \left[1+ \dot{m}(t)\right] \left[1+\varepsilon_1 q(t)\right] -1,
\end{equation}
where $q(t)$ is the QPO light-curve and constant $\varepsilon_1$ regulates its prominence (subscript 1 corresponds to the first harmonic).
Such light-curve can arise, e.g., in the model where QPOs are produced by the rotating emission pattern of the hot flow \citep{VPI13}, so the QPO 
modulates the mass accretion rate fluctuations in a multiplicative way. 
The most common scenario having the entire hot flow rotating is when it undergoes the Lense-Thirring precession \citep{FB07,IDF09}.

As an illustration, we consider two limiting cases where the QPO central frequency coincides with the peak and with the dip of the PSD.
The parameters are: $f_{\rm QPO}=0.7$~Hz and $0.5$~Hz (respectively); $\varepsilon_1=1$ and $\Delta f_{\rm QPO}=0.05$~Hz.
We obtain the X-ray light-curves using Equation~(\ref{eq:x_qpo}). 
To reduce the noise appearing due to coupling of QPO and broadband variability, we simulate 1000 light-curves, calculate their PSDs and average over the realizations.
The resulting PSDs are shown in Figure~\ref{fig:qpo}.
For a comparison, we also plot the model with the same parameters but without QPO (black line, see parameters in Table~\ref{tab:par}).
We see that the presence of the QPO reduces the oscillation prominence at high frequencies.

The described model with the QPO requires numerical modelling, which may be computationally expensive when trying to fit the data directly.
A simplest way to introduce the QPO into the model is just to add the QPO narrow Lorentzian to the model PSD.
However, this procedure implicitly assumes that the QPO is related to the broadband variability in an additive way.
We demonstrate the difference between multiplicative and additive QPO coupling in Figure~\ref{fig:qpo}.
Increasing $\varepsilon_{\rm ds}$, i.e. suppressing the oscillations in the additive model, does not eliminate the difference between the two models: when the consistence is 
achieved at frequencies $f<f_{\rm QPO}$, the oscillations at frequencies $f>f_{\rm QPO}$ remain more pronounced in the additive model.
Hence, the total PSD fitted using the analytical model and a QPO Lorentzian remains somewhat different from the one obtained using multiplicative model.

\section{Implications}

\subsection{Observational appearance}\label{sect:obs_appear}

Multiple peaks in the X-ray PSDs were reported, e.g., in GX~339--4 \citep{Belloni06,DeMarco15}, Cyg X-1 \citep{GCR99}, XTE~J1748--288 \citep{RTB00}, 
SWIFT~J1753.5--0127 \citep{HBM09},  XTE J1550--564 \citep{Homan01}, GS~1124--68 \citep{vdK95}.
The peaks can be fitted with a number of Lorentzians (zero-centered and QPO-like features), but the physical explanation behind this description is still missing.
The same data can be explained in terms of the proposed model.
We compute the PSD assuming the QPO is coupled to the broadband noise in an additive way, so the model PSD is given by 
\begin{equation}
  |X|^2(f)=r \left[ |X|_{\rm bb}^2  + \varepsilon_{n}{\cal Q}_n(f) \right], \; n=1,2
\end{equation}
where the broadband variability $|X|_{\rm bb}^2$ is described by Equation~(\ref{eq:xpsd_cross}) and the two QPO harmonics are described by narrow 
Lorentzians ${\cal Q}_n(f)$ (Equation~\ref{eq:qpo_lor}).
Parameters $\varepsilon_n$ describe relative importance of the QPO harmonics (first and second) and $r$ is the overall normalization. 
We take the mass accretion rate PSD in the form
\begin{equation}
  |\dot{M}|^2(f) = \frac {f^{\gamma_1}}{1+(f/f_1)^{\gamma_2}+(f/f_2)^{\gamma_3}}
\end{equation}
to mimic the double-broken power-law ($f_1$ and $f_2$ are the two break frequencies).
We expect the indices to be $\gamma_1\sim0$ and $\gamma_2\sim1$, as in the model of propagation fluctuations.

\begin{figure}
\figurenum{5}
\plotone{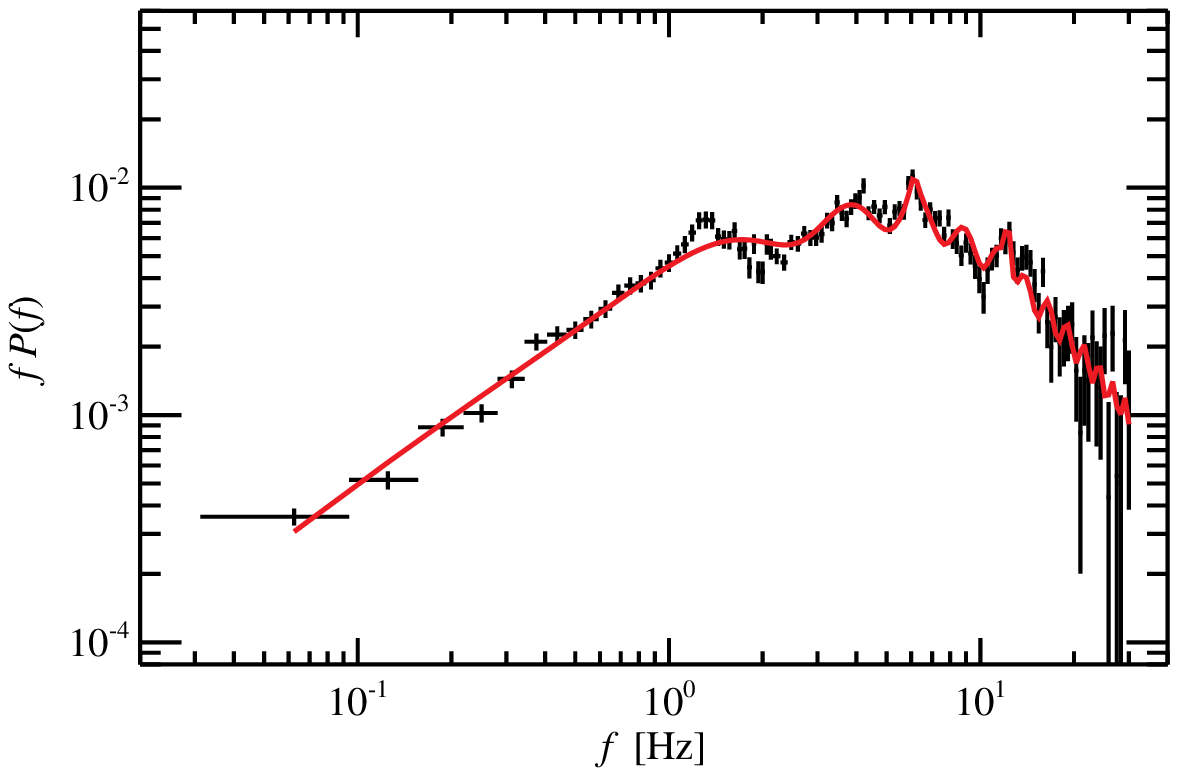}
\caption{
PSD simulated from the model (red solid line) resembles the observed in GX~339--4 \citep[black crosses,][]{Belloni06}.
The model parameters are listed in Table~\ref{tab:par_real}.
}\label{fig:real} 
\end{figure}

As the first example, we apply the developed model to the 2--15~keV range PSD of black hole binary GX~339--4 (see Figure~\ref{fig:real}).
The data were taken during the hard-intermediate state and reported in \citet{Belloni06}. 
The parameters of our modeling are listed in Table~\ref{tab:par_real}.
We note that the delay and sign of $\varepsilon_{\rm ds}$ can be easily deduced from the characteristic frequencies of the peaks that appear at approximately 
1.3, 4, 6, 9 and 13~Hz.
The third peak is likely affected by the QPO, however, even from this sequence the 1:3:5:7:9 ratio can be recovered.
From this we deduce that the disk and synchrotron Comptonization light-curves are anti-correlated and the delay time is $t_0=0.38$~s, close to the model value.

\begin{figure}
\figurenum{6}
\plotone{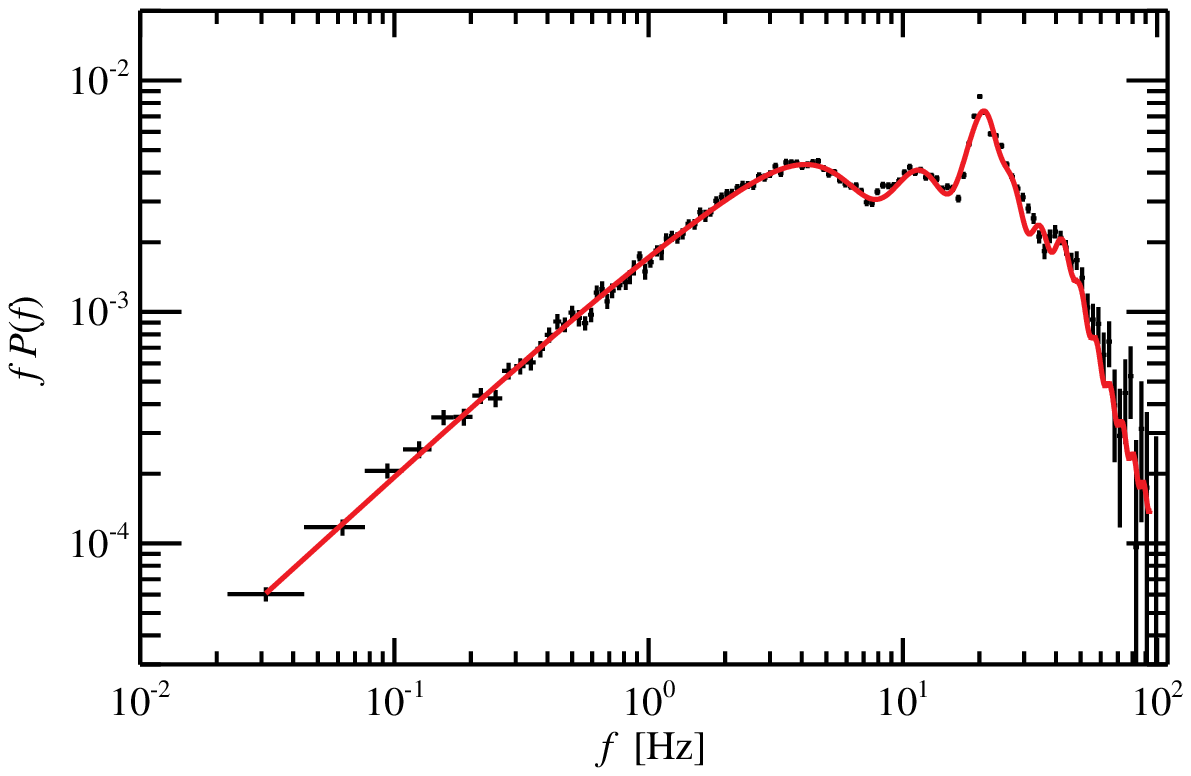}
\caption{
PSD simulated from the model (red solid line) resembles the observed in XTE~J1748--288 \citep[black crosses,][]{RTB00}.
The model parameters are listed in Table~\ref{tab:par_real}.
}\label{fig:real2} 
\end{figure}

Another example of model application is the hard-intermediate state of black hole binary XTE~J1748--288 \citep{RTB00}.
We apply the same additive model and show the fit in Figure~\ref{fig:real2}, parameters are listed in Table~\ref{tab:par_real}.
Again, the $t_0$ value can be recovered from the peak positions. 

Comparing the fitting parameters for these two objects we notice that the low-frequency break $f_1$ in the data on XTE~J1748--288 is three times larger than 
in the data on GX~339--4, by the same factor as the ratio of the inferred delay times in these objects.
This is in agreement with the picture that both the delay time and the low-frequency break of the PSD are related to the disk truncation radius.
In addition, the disk fraction $\varepsilon_{\rm ds}$ is somewhat higher for XTE~J1748--288, again consistent with the disk being closer to the BH in this object.
The QPO frequency, yet another marker of the disk truncation radius, is higher in XTE~J1748--288, again suggesting smaller truncation radius here.

When the disk is close to the precessing hot flow, its emission becomes modulated at the QPO frequency, so we have two sources of the QPO: 
one from the inner hot accretion flow, and the other from the modulation of the soft disk photons.
Interference of these QPOs may potentially explain the type-B QPOs that also appear at the intermediate states.
Interestingly, a PSD with a prominent type-B QPO and the PSD showed in Figure~\ref{fig:real} were detected one hour apart \citep[see][]{Belloni06}.

\begin{table}
\caption{Model parameters for Figures~\ref{fig:real}~and~\ref{fig:real2}.
}\label{tab:par_real}
  \begin{center}
\begin{tabular}{ccc}
\hline
\hline
{Parameter}			& {GX~339--4}		& {XTE~J1748--288}		\\
\hline 
$r$					&  $6.78\times10^{-3}$& $2.22\times10^{-3}$		\\
$f_1$~(Hz)			&  1.3			& 2.88		\\
$f_2$~(Hz)			&  5.0			& 20.1		\\
$\gamma_1$			&  0.077			& 0.007		\\
$\gamma_2$			&  0.96			& 1.45		\\
$\gamma_3$			& 3.14			& 5.11		\\
$\varepsilon_{\rm ds}$	& $-$0.068		& $-$0.086		\\
 $t_0$~(s)				& 0.39			& 0.13	\\
 $f_{\rm filt}$~(Hz)		& $\infty$			& $\infty$	\\
 $f_{\rm QPO}$~(Hz)		& 6.1				& 21 	\\
 $\Delta f_{\rm QPO, 1}$~(Hz)& 0.34 			& 3.0	\\
 $\varepsilon_1$		& 0.08			& 1.0	\\
$\Delta f_{\rm QPO, 2}$~(Hz)& 0.32			& 7.2 	\\
 $\varepsilon_2$		& 0.039			& 0.14	\\
\hline
     \end{tabular}
  \end{center}
\end{table}

\subsection{Viscous timescale marker}

The proposed model provides a potential to explore viscous timescale.
The time-lags between two X-ray components can be attributed to the propagation time between two regions of Comptonization.
The disk Comptonization likely occurs close to its truncation radius and the synchrotron Comptonization operates in the vicinity of the black hole.
Hence, we can roughly estimate the propagation time with the viscous timescale at the truncation radius. 
The latter is determined as $1/t_{\rm visc}=\Omega_{\rm K} \alpha (H/R)^2$, where $\Omega_{\rm K}$ is the Keplerian angular velocity, $\alpha$ is the viscosity 
parameter and $H/R$ is the disk height-to-radius ratio.
Scaling the radii to $50R_{\rm S}$ ($R_{\rm S}$ is the Schwarzschild radius), the typical truncation radius in the hard state \citep{GCR00,BZ16}, we obtain
\begin{equation}
  \alpha \left( \frac{H}{R} \right)^2 = \frac{0.05}{t_{\rm visc}} \left( \frac{M_{\rm BH}}{10 M_{\odot}}\right) \left( \frac{R}{50R_{\rm S}}\right)^{3/2},
\end{equation}
where $M_{\rm BH}$ is the black hole mass.
The disk truncation radius during the intermediate state is likely smaller.
It cannot be very large, as it should contribute substantial amount of soft photons for Comptonization, but cannot be too small either, as it would 
produce too soft spectrum.

We impose constraints on $\alpha$ and $H/R$ by considering the range of possible truncation radii, $10-50R_{\rm S}$, for $H/R<1$, and the condition of appearance of the 
QPO (from the solid-body precession of the hot disc, \citealt{FB07}) $\alpha<H/R$.
These limitations lead to the ranges $0.23<H/R$ and $0.01<\alpha<0.5$ for GX~339--4 and $0.33<H/R$ and $0.03<\alpha<0.73$ for XTE~J1748--288.
Constraints on $H/R$ suggest that the propagation occurs in the geometrically thick flow in both cases.
This implies  that we are probing here the viscous parameter of the hot accretion flow, not that of the standard \citet{SS73} accretion disc, 
which is measured in cataclysmic variables, $\alpha_{\rm CV}=0.2$ \citep{Smak99alpha}.
The viscosity parameter of an advection dominated accretion flow is related to the maximal luminosity at which it can exist as 
$L_{\rm max}\sim0.4\alpha^2L_{\rm Edd}$ 
(where $L_{\rm Edd}$ is the Eddington luminosity), whereas for the luminous hot accretion flow, where the electron cooling is significant, 
the maximal luminosity is $\sim\alpha^2 L_{\rm Edd}$ \citep{YN14}. 
The latter hot flow model produces spectra resembling those of the hard-state black holes and Seyfert galaxies \citep{yuanaaz04,YZ07}.
For $\alpha=0.5$ we obtain $L_{\rm max}=0.1L_{\rm Edd}$ and $0.25L_{\rm Edd}$ for advection-dominated and luminous hot accretion flow, respectively.

\subsection{Agreement with other observables}

The proposed model assumes that the X-rays are produced by two processes, the disk and synchrotron Comptonization, which are drastically different in terms of 
spectral behavior.
We hence should be able to detect distinctive signatures of those in the spectral data.
Investigations of the dependence of the X-ray spectral slope on luminosity show that at high luminosity the spectral index correlates with 
$L/L_{\rm Edd}$, but when this ratio falls below $\sim$0.01, the dependence is reversed, and for lower $L/L_{\rm Edd}$ the index is larger \citep{SPDM11,YXYZ15}.
The behavior is naturally explained in terms of change of seed photon source, the disk being at high $L/L_{\rm Edd}$ and 
the synchrotron contributing at small $L/L_{\rm Edd}$ \citep[figure~7b of][]{VVP11}.
Recently, an origin of seed photons for Comptonization was investigated using the data on the outburst decline of black hole binary SWIFT~J1753.5--0127 \citep{KVT16}.
The dependence of spectral index on the ratio of seed to Comptonized luminosity followed two different tracks at the peak and at the tail of the outburst. 
This was interpreted as the change of seed photon source, from the cold disk at the outburst peak to the synchrotron emission at the outburst tail.
There should exist an overlap, where both mechanisms effectively operate, which likely corresponds to an intermediate state.
Interestingly, this object was observed simultaneously in the X-rays and optical during the decline phase \citep{HBM09}.
The X-ray PSD demonstrates the peaked noise component, and the optical/X-ray cross-correlation function shows multiple dips and peaks shape,
which can only be explained by considering two X-ray components (Veledina et al. in prep.).

In terms of the proposed model, the intermediate state corresponds to $\varepsilon_{\rm ds}\sim1$, the hard state can be described by $\varepsilon_{\rm ds}\ll1$ and 
the soft state, when there still exists a hot medium above the disc, can be described by $\varepsilon_{\rm ds}\gg1$.
The cross-term becomes more pronounced when the contributions of two terms are equal, thus in the low-luminosity hard state and very close to the soft state we do not 
expect multiple peaks.
This general expectation is in agreement with the patterns seen in Cyg~X-1 \citep{GPB14}: the two-peak structure is most prominent at the intermediate spectral states.

We note that the cross-term only redistributes the overall power, but does not lead to a considerable suppression/enhancement of the 
root mean square (rms) variability amplitude, this can be seen by integrating Equation~(\ref{eq:xpsd_cross}) over all frequencies.
This implies that the presence of an additional component does not alter the rms-flux relation \citep{UM01}, which is an essential piece of evidence in favor of 
propagating accretion fluctuations model.

\citet{GNC08} found that the high-frequency part of PSD, above the high-frequency break (if we utilize the double-broken power-law model), remains remarkably 
similar in a number of black holes.
The slope and normalization constant are the same, in any particular object, during the state transition.
They attributed this high-frequency part to a signature of the last stable orbit.
We note that in our model, the slope of high-frequency part is not altered under change of any parameter.
Resulting PSD at high $f$ has the same average slope as the seed Lorentzian has (apart from possible oscillations around the mean).

Recent investigations of variability at energies below 2~keV with the {\it XMM}--Newton satellite suggest the variability is enhanced, compared to hard X-rays, 
at long timescales \citep{WU09,UWC11,CUM12}.
In the proposed model, probing different energy spectral ranges can be thought in terms of changing $\varepsilon_{\rm ds}$ parameter.
Slight increase of power at low  frequencies ($\sim0.1$~Hz in the present setup, but this depends on the seed Lorentzian parameters) can be obtained if relative 
contribution of disk Comptonization $\varepsilon_{\rm ds}=1$ for energies below 2~keV and $\varepsilon_{\rm ds}=0.3$ for energies above 2~keV 
(Figure~\ref{fig:model_params}b, red and blue lines, respectively).
There also can be some difference in delay time, as the softer synchrotron Comptonization spectra are produced further away from the BH, 
i.e. closer to truncation radius \citep{VPV13}.
Thus, for soft X-rays the delay is less than for hard X-rays, hence from Figure~\ref{fig:model_params}a we deduce that the soft X-rays have larger variability amplitude at 
frequencies $\sim$0.01~Hz than hard X-rays.

Another intriguing behavior is demonstrated by the time-lag spectra, which show the broken power-law dependence on energy in the low-frequency range: 
a weak increase of time-lag with energy above $\sim2$~keV is opposed to the strong dependence on energy below this energy \citep{WU09,UWC11}.
In addition, the dependence of time-lags on Fourier frequency depends on the energy range: it is a simple power-law for hard-medium energies, but the  
medium-soft time-lags demonstrate a steeper slope at high frequencies, and the overall shape is a broken power-law.
These features can be reproduced if we consider two components, one dominating in the soft energy range and the other is dominating at higher energies, which is delayed 
with respect to the soft one.
We note that in this case the time-lag corresponds to viscous timescale, not to the light-travel time.
Rapid increase of the time-lags at softer energies suggests a steep soft-component spectrum, it is not clear whether the disk Comptonization spectrum can account for that
if we assume it is variable with constant spectral shape.
Spectral evolution might play an important role here, e.g. it was shown that spectral pivoting can produce the broken power-law dependence of time-lags \citep{PF99}.

Recently, two distinct classes of time-lags between the soft (presumably coming from the disc) and the hard (coming from the innermost parts of accretion flow) 
X-rays were identified \citep{DPM15}.
The lags measured at low Fourier frequencies are of the order of 0.1~s and are likely connected to the propagation times from the cold disc to the Comptonization region.
The measured time-lags are in agreement with the delays ($t_0$) obtained in this paper.
The reverberation lags constitute another class, in which case the soft band is delayed with respect to the hard band by about $10^{-3}-10^{-2}$~s.
Such echoes appear when the X-rays produced intrinsically within the flow are reflected from the cold accretion disc \citep[see, e.g.][]{GCR00,Pou02}, 
and the X-ray PSD in this case is expected to have similar bumps due to the interference of the primary and reflected emission \citep{PPD16}.
However, due to substantial smearing of the reflected component, the bumps are likely less prominent than those proposed in this work.

An independent way to measure the viscosity parameter is from the low-frequency break of the X-ray power spectrum.
The characteristic increase of the break frequency observed at the hard-to-soft state transition is likely caused by the decrease of the disc truncation radius \citep{GCR99}.
Assuming the break frequency corresponds to the viscous timescale at the disc truncation radius, the viscosity and the disc aspect ratio $H/R$ can be recovered.
Evolution of the $H/R$ throughout the outburst of BH binary XTE~J1550--564 was traced by fitting to the low-frequency break of the X-ray PSD \citep{ID12}.
Such modeling relies on the prescribed distribution for the hot flow surface density distribution, so it would be interesting to compare the $H/R$ obtained by the two methods.

The proposed model can also be applied to other accreting compact objects, such as neutron star X-ray binaries and active galactic nuclei.
Neutron star binaries reveal bumps (peaked noise) in their spectra \citep[e.g.,][]{HvdK89,BH90atlas,vdK06}.
They have an additional X-ray component, the boundary layer, which can also interfere with the two considered components.
Measuring the PSDs in active galactic nuclei is difficult due to an intermittent sampling of the frequency range of interest \citep{UM05}.
The available data is generally consistent with a power-law or a broken power-law \citep{EN99,UMP02}, however, 
humps are sometimes observed (most prominent are in PKS~1346+26, some are seen in NGC~6860, a dip is present in 3C59, \citealt{GMV12}; see also \citealt{EPE16}).
Variability in active galactic nuclei can also studied using the structure function, which is expected to demonstrate a dip at the delay time.

\section{Conclusions}\label{sect:conclus}

Power spectra of accreting black hole binaries only roughly resemble flat-top noise (double-broken power-law).
Instead, the data are more consistent with being produced by two or more components, usually fitted with Lorentzians. 
In some cases, up to six Lorentzians were required to explain its complex shape.
We propose a physical scenario explaining appearance of such peaked noise in power spectra of accreting black hole binaries.

We consider the X-rays arising from two Comptonization processes: in up-scattering of the disk and the synchrotron photons.
The two components are assumed to linearly respond to the mass accretion rate fluctuations, however, there is a time delay between them, corresponding to the 
propagation timescale between the two radii where Comptonization (of disk and of synchrotron photons) occurs.
The total X-ray light-curve is obtained as a sum of two, time-shifted, mass accretion rate light-curves, and the power spectral density of such process constitutes a 
number of humps, appearing from the cross-term.
We show that the humps central frequencies are harmonically related and the dips of power spectra follow 1:3:5 relation if the disk and synchrotron terms are correlated.
If the two processes are anti-correlated, then the peaks follow 1:3:5 relation and the dips are harmonically related.
We also show that coupling with the QPO reduces the cross-term amplitude, so the humps in the power spectra are less pronounced.

We discuss the consistency of this scenario with a number of observables such as: correlation of appearance of spectral bumps with 
the spectral properties, common shape of the power spectra at high frequencies (above the high-frequency break in terms of double-broken power-law model),
difference in power observed for soft and hard energies.
We directly compare the model to the power spectra of black hole binaries GX~339--4 and XTE~J1748--288 demonstrating numerous peaks.
From the inferred time delay we put constraints on a combination of the viscosity parameter and the disk height-to-radius ratio $\alpha (H/R)^2$ and deduce that  
the propagation occurs in geometrically thick flow, $H/R>0.2$, and $\alpha<0.7$ for this hot flow.
The model can be extended to describe power spectra of neutron stars and active galactic nuclei.

\section*{Acknowledgements}

The author is grateful to Juri Poutanen and Andrzej Zdziarski for discussions and comments on the manuscript and Mike Revnivtsev and Tomaso Belloni for the data.
I thank the referee for the suggestions which improved the paper.
The work was supported by the Academy of Finland grant 268740.


\begin{thebibliography}{}
\expandafter\ifx\csname natexlab\endcsname\relax\def\natexlab#1{#1}\fi

\bibitem[{{Ar{\'e}valo} \& {Uttley}(2006)}]{AU06}
{Ar{\'e}valo}, P., \& {Uttley}, P. 2006, \mnras, 367, 801

\bibitem[{{Axelsson} {et~al.}(2005){Axelsson}, {Borgonovo}, \&
  {Larsson}}]{ABL05}
{Axelsson}, M., {Borgonovo}, L., \& {Larsson}, S. 2005, \aap, 438, 999

\bibitem[{{Basak} \& {Zdziarski}(2016)}]{BZ16}
{Basak}, R., \& {Zdziarski}, A.~A. 2016, \mnras, 458, 2199

\bibitem[{{Belloni} \& {Hasinger}(1990{\natexlab{a}})}]{BH90atlas}
{Belloni}, T., \& {Hasinger}, G. 1990{\natexlab{a}}, \aap, 230, 103

\bibitem[{{Belloni} \& {Hasinger}(1990{\natexlab{b}})}]{BH90}
---. 1990{\natexlab{b}}, \aap, 227, L33

\bibitem[{{Belloni} {et~al.}(2002){Belloni}, {Psaltis}, \& {van der
  Klis}}]{BPvdK02}
{Belloni}, T., {Psaltis}, D., \& {van der Klis}, M. 2002, \apj, 572, 392

\bibitem[{{Belloni} {et~al.}(2006){Belloni}, {Parolin}, {Del Santo}, {Homan},
  {Casella}, {Fender}, {Lewin}, {M{\'e}ndez}, {Miller}, \& {van der
  Klis}}]{Belloni06}
{Belloni}, T., {Parolin}, I., {Del Santo}, M., {et~al.} 2006, \mnras, 367, 1113

\bibitem[{{Casella} {et~al.}(2005){Casella}, {Belloni}, \& {Stella}}]{CBS05}
{Casella}, P., {Belloni}, T., \& {Stella}, L. 2005, \apj, 629, 403

\bibitem[{{Cassatella} {et~al.}(2012){Cassatella}, {Uttley}, \&
  {Maccarone}}]{CUM12}
{Cassatella}, P., {Uttley}, P., \& {Maccarone}, T.~J. 2012, \mnras, 427, 2985

\bibitem[{{Churazov} {et~al.}(2001){Churazov}, {Gilfanov}, \&
  {Revnivtsev}}]{CGR01}
{Churazov}, E., {Gilfanov}, M., \& {Revnivtsev}, M. 2001, \mnras, 321, 759

\bibitem[{{De Marco} {et~al.}(2015{\natexlab{a}}){De Marco}, {Ponti},
  {Mu{\~n}oz-Darias}, \& {Nandra}}]{DeMarco15}
{De Marco}, B., {Ponti}, G., {Mu{\~n}oz-Darias}, T., \& {Nandra}, K.
  2015{\natexlab{a}}, \mnras, 454, 2360

\bibitem[{{De Marco} {et~al.}(2015{\natexlab{b}}){De Marco}, {Ponti},
  {Mu{\~n}oz-Darias}, \& {Nandra}}]{DPM15}
---. 2015{\natexlab{b}}, \apj, 814, 50

\bibitem[{{Done} {et~al.}(2007){Done}, {Gierli{\'n}ski}, \& {Kubota}}]{DGK07}
{Done}, C., {Gierli{\'n}ski}, M., \& {Kubota}, A. 2007, \aapr, 15, 1

\bibitem[{{Edelson} \& {Nandra}(1999)}]{EN99}
{Edelson}, R., \& {Nandra}, K. 1999, \apj, 514, 682

\bibitem[{{Emmanoulopoulos} {et~al.}(2016){Emmanoulopoulos}, {Papadakis},
  {Epitropakis}, {Pech{\'a}{\v c}ek}, {Dov{\v c}iak}, \& {McHardy}}]{EPE16}
{Emmanoulopoulos}, D., {Papadakis}, I.~E., {Epitropakis}, A., {et~al.} 2016,
  \mnras, 461, 1642

\bibitem[{{Esin} {et~al.}(1997){Esin}, {McClintock}, \& {Narayan}}]{Esin97}
{Esin}, A.~A., {McClintock}, J.~E., \& {Narayan}, R. 1997, \apj, 489, 865

\bibitem[{{Fragile}(2009)}]{F09}
{Fragile}, P.~C. 2009, \apjl, 706, L246

\bibitem[{{Fragile} {et~al.}(2007){Fragile}, {Blaes}, {Anninos}, \&
  {Salmonson}}]{FB07}
{Fragile}, P.~C., {Blaes}, O.~M., {Anninos}, P., \& {Salmonson}, J.~D. 2007,
  \apj, 668, 417

\bibitem[{{Gandhi} {et~al.}(2008){Gandhi}, {Makishima}, {Durant}, {Fabian},
  {Dhillon}, {Marsh}, {Miller}, {Shahbaz}, \& {Spruit}}]{GMD08}
{Gandhi}, P., {Makishima}, K., {Durant}, M., {et~al.} 2008, \mnras, 390, L29

\bibitem[{{Gierli{\'n}ski} {et~al.}(2008){Gierli{\'n}ski}, {Niko{\l}ajuk}, \&
  {Czerny}}]{GNC08}
{Gierli{\'n}ski}, M., {Niko{\l}ajuk}, M., \& {Czerny}, B. 2008, \mnras, 383,
  741

\bibitem[{{Gilfanov} {et~al.}(1999){Gilfanov}, {Churazov}, \&
  {Revnivtsev}}]{GCR99}
{Gilfanov}, M., {Churazov}, E., \& {Revnivtsev}, M. 1999, \aap, 352, 182

\bibitem[{{Gilfanov} {et~al.}(2000){Gilfanov}, {Churazov}, \&
  {Revnivtsev}}]{GCR00}
---. 2000, \mnras, 316, 923

\bibitem[{{Gonz{\'a}lez-Mart{\'{\i}}n} \& {Vaughan}(2012)}]{GMV12}
{Gonz{\'a}lez-Mart{\'{\i}}n}, O., \& {Vaughan}, S. 2012, \aap, 544, A80

\bibitem[{{Grinberg} {et~al.}(2014){Grinberg}, {Pottschmidt}, {B{\"o}ck},
  {Schmid}, {Nowak}, {Uttley}, {Tomsick}, {Rodriguez}, {Hell}, {Markowitz},
  {Bodaghee}, {Cadolle Bel}, {Rothschild}, \& {Wilms}}]{GPB14}
{Grinberg}, V., {Pottschmidt}, K., {B{\"o}ck}, M., {et~al.} 2014, \aap, 565, A1

\bibitem[{{Hasinger} \& {van der Klis}(1989)}]{HvdK89}
{Hasinger}, G., \& {van der Klis}, M. 1989, \aap, 225, 79

\bibitem[{{Hogg} \& {Reynolds}(2015)}]{HR15}
{Hogg}, J.~D., \& {Reynolds}, C. 2015, ArXiv e-prints, arXiv:1512.05350

\bibitem[{{Homan} {et~al.}(2001){Homan}, {Wijnands}, {van der Klis}, {Belloni},
  {van Paradijs}, {Klein-Wolt}, {Fender}, \& {M{\'e}ndez}}]{Homan01}
{Homan}, J., {Wijnands}, R., {van der Klis}, M., {et~al.} 2001, \apjs, 132, 377

\bibitem[{{Hynes} {et~al.}(2009){Hynes}, {O'Brien}, {Mullally}, \&
  {Ashcraft}}]{HBM09}
{Hynes}, R.~I., {O'Brien}, K., {Mullally}, F., \& {Ashcraft}, T. 2009, \mnras,
  399, 281

\bibitem[{{Ingram} \& {Done}(2011)}]{ID11}
{Ingram}, A., \& {Done}, C. 2011, \mnras, 415, 2323

\bibitem[{{Ingram} \& {Done}(2012)}]{ID12}
---. 2012, \mnras, 419, 2369

\bibitem[{{Ingram} {et~al.}(2009){Ingram}, {Done}, \& {Fragile}}]{IDF09}
{Ingram}, A., {Done}, C., \& {Fragile}, P.~C. 2009, \mnras, 397, L101

\bibitem[{{Ingram} \& {van der Klis}(2013)}]{IvdK13}
{Ingram}, A., \& {van der Klis}, M. 2013, \mnras, 434, 1476

\bibitem[{{Kajava} {et~al.}(2016){Kajava}, {Veledina}, {Tsygankov}, \&
  {Neustroev}}]{KVT16}
{Kajava}, J.~J.~E., {Veledina}, A., {Tsygankov}, S., \& {Neustroev}, V. 2016,
  ArXiv e-prints, arXiv:1603.08796

\bibitem[{{Kotov} {et~al.}(2001){Kotov}, {Churazov}, \& {Gilfanov}}]{KCG01}
{Kotov}, O., {Churazov}, E., \& {Gilfanov}, M. 2001, \mnras, 327, 799

\bibitem[{{Lyubarskii}(1997)}]{Lyub97}
{Lyubarskii}, Y.~E. 1997, \mnras, 292, 679

\bibitem[{{Malzac} {et~al.}(2003){Malzac}, {Belloni}, {Spruit}, \&
  {Kanbach}}]{MBSK03}
{Malzac}, J., {Belloni}, T., {Spruit}, H.~C., \& {Kanbach}, G. 2003, \aap, 407,
  335

\bibitem[{{Malzac} \& {Belmont}(2009)}]{MB09}
{Malzac}, J., \& {Belmont}, R. 2009, \mnras, 392, 570

\bibitem[{{Nolan} {et~al.}(1981){Nolan}, {Gruber}, {Matteson}, {Peterson},
  {Rothschild}, {Doty}, {Levine}, {Lewin}, \& {Primini}}]{NGM81}
{Nolan}, P.~L., {Gruber}, D.~E., {Matteson}, J.~L., {et~al.} 1981, \apj, 246,
  494

\bibitem[{{Nowak} {et~al.}(1999){Nowak}, {Wilms}, \& {Dove}}]{NWD99}
{Nowak}, M.~A., {Wilms}, J., \& {Dove}, J.~B. 1999, \apj, 517, 355

\bibitem[{{Papadakis} {et~al.}(2016){Papadakis}, {Pech{\'a}{\v c}ek}, {Dov{\v
  c}iak}, {Epitropakis}, {Emmanoulopoulos}, \& {Karas}}]{PPD16}
{Papadakis}, I., {Pech{\'a}{\v c}ek}, T., {Dov{\v c}iak}, M., {et~al.} 2016,
  \aap, 588, A13

\bibitem[{{Pottschmidt} {et~al.}(2003){Pottschmidt}, {Wilms}, {Nowak},
  {Pooley}, {Gleissner}, {Heindl}, {Smith}, {Remillard}, \& {Staubert}}]{PWN03}
{Pottschmidt}, K., {Wilms}, J., {Nowak}, M.~A., {et~al.} 2003, \aap, 407, 1039

\bibitem[{{Poutanen}(2002)}]{Pou02}
{Poutanen}, J. 2002, \mnras, 332, 257

\bibitem[{{Poutanen} \& {Fabian}(1999)}]{PF99}
{Poutanen}, J., \& {Fabian}, A.~C. 1999, \mnras, 306, L31

\bibitem[{{Poutanen} {et~al.}(1997){Poutanen}, {Krolik}, \& {Ryde}}]{PKR97}
{Poutanen}, J., {Krolik}, J.~H., \& {Ryde}, F. 1997, \mnras, 292, L21

\bibitem[{{Poutanen} \& {Veledina}(2014)}]{PV14}
{Poutanen}, J., \& {Veledina}, A. 2014, \ssr, 183, 61

\bibitem[{{Poutanen} \& {Vurm}(2009)}]{PV09}
{Poutanen}, J., \& {Vurm}, I. 2009, \apjl, 690, L97

\bibitem[{{Revnivtsev} {et~al.}(2000){Revnivtsev}, {Trudolyubov}, \&
  {Borozdin}}]{RTB00}
{Revnivtsev}, M.~G., {Trudolyubov}, S.~P., \& {Borozdin}, K.~N. 2000, \mnras,
  312, 151

\bibitem[{{Shakura} \& {Sunyaev}(1973)}]{SS73}
{Shakura}, N.~I., \& {Sunyaev}, R.~A. 1973, \aap, 24, 337

\bibitem[{{Smak}(1999)}]{Smak99alpha}
{Smak}, J. 1999, \actaa, 49, 391

\bibitem[{{Sobolewska} {et~al.}(2011){Sobolewska}, {Papadakis}, {Done}, \&
  {Malzac}}]{SPDM11}
{Sobolewska}, M.~A., {Papadakis}, I.~E., {Done}, C., \& {Malzac}, J. 2011,
  \mnras, 417, 280

\bibitem[{{Terrell}(1972)}]{Terrell72}
{Terrell}, Jr., N.~J. 1972, \apjl, 174, L35

\bibitem[{{Timmer} \& {Koenig}(1995)}]{TK95}
{Timmer}, J., \& {Koenig}, M. 1995, \aap, 300, 707

\bibitem[{{Uttley} \& {McHardy}(2001)}]{UM01}
{Uttley}, P., \& {McHardy}, I.~M. 2001, \mnras, 323, L26

\bibitem[{{Uttley} \& {McHardy}(2005)}]{UM05}
---. 2005, \mnras, 363, 586

\bibitem[{{Uttley} {et~al.}(2002){Uttley}, {McHardy}, \& {Papadakis}}]{UMP02}
{Uttley}, P., {McHardy}, I.~M., \& {Papadakis}, I.~E. 2002, \mnras, 332, 231

\bibitem[{{Uttley} {et~al.}(2011){Uttley}, {Wilkinson}, {Cassatella}, {Wilms},
  {Pottschmidt}, {Hanke}, \& {B{\"o}ck}}]{UWC11}
{Uttley}, P., {Wilkinson}, T., {Cassatella}, P., {et~al.} 2011, \mnras, 414,
  L60

\bibitem[{{van der Klis}(1995)}]{vdK95}
{van der Klis}, M. 1995, in Frontier Objects in Astrophysics and Particle
  Physics, ed. F.~{Giovannelli} \& G.~{Mannocchi}, 213

\bibitem[{{van der Klis}(2006)}]{vdK06}
{van der Klis}, M. 2006, in Compact stellar X-ray sources, Cambridge
  Astrophysics Series, No. 39, ed. W.~{Lewin} \& M.~{van der Klis} (Cambridge:
  Cambridge University Press), 39--112

\bibitem[{{Veledina} {et~al.}(2013{\natexlab{a}}){Veledina}, {Poutanen}, \&
  {Ingram}}]{VPI13}
{Veledina}, A., {Poutanen}, J., \& {Ingram}, A. 2013{\natexlab{a}}, \apj, 778,
  165

\bibitem[{{Veledina} {et~al.}(2011{\natexlab{a}}){Veledina}, {Poutanen}, \&
  {Vurm}}]{VPV11}
{Veledina}, A., {Poutanen}, J., \& {Vurm}, I. 2011{\natexlab{a}}, \apjl, 737,
  L17

\bibitem[{{Veledina} {et~al.}(2013{\natexlab{b}}){Veledina}, {Poutanen}, \&
  {Vurm}}]{VPV13}
---. 2013{\natexlab{b}}, \mnras, 430, 3196

\bibitem[{{Veledina} {et~al.}(2011{\natexlab{b}}){Veledina}, {Vurm}, \&
  {Poutanen}}]{VVP11}
{Veledina}, A., {Vurm}, I., \& {Poutanen}, J. 2011{\natexlab{b}}, \mnras, 414,
  3330

\bibitem[{{Vikhlinin} {et~al.}(1994){Vikhlinin}, {Churazov}, \&
  {Gilfanov}}]{VCG94b}
{Vikhlinin}, A., {Churazov}, E., \& {Gilfanov}, M. 1994, \aap, 287, 73

\bibitem[{{Wilkinson} \& {Uttley}(2009)}]{WU09}
{Wilkinson}, T., \& {Uttley}, P. 2009, \mnras, 397, 666

\bibitem[{{Yang} {et~al.}(2015){Yang}, {Xie}, {Yuan}, {Zdziarski},
  {Gierli{\'n}ski}, {Ho}, \& {Yu}}]{YXYZ15}
{Yang}, Q.-X., {Xie}, F.-G., {Yuan}, F., {et~al.} 2015, \mnras, 447, 1692

\bibitem[{{Yuan} \& {Narayan}(2014)}]{YN14}
{Yuan}, F., \& {Narayan}, R. 2014, \araa, 52, 529

\bibitem[{{Yuan} \& {Zdziarski}(2004)}]{yuanaaz04}
{Yuan}, F., \& {Zdziarski}, A.~A. 2004, \mnras, 354, 953

\bibitem[{{Yuan} {et~al.}(2007){Yuan}, {Zdziarski}, {Xue}, \& {Wu}}]{YZ07}
{Yuan}, F., {Zdziarski}, A.~A., {Xue}, Y., \& {Wu}, X.-B. 2007, \apj, 659, 541

\bibitem[{{Zdziarski} \& {Gierli{\'n}ski}(2004)}]{ZG04}
{Zdziarski}, A.~A., \& {Gierli{\'n}ski}, M. 2004, Progr. Theor. Phys. Suppl.,
  155, 99

\bibitem[{{Zdziarski} {et~al.}(2004){Zdziarski}, {Gierli{\'n}ski},
  {Miko{\l}ajewska}, {Wardzi{\'n}ski}, {Smith}, {Harmon}, \&
  {Kitamoto}}]{ZGM04}
{Zdziarski}, A.~A., {Gierli{\'n}ski}, M., {Miko{\l}ajewska}, J., {et~al.} 2004,
  \mnras, 351, 791

\bibitem[{{Zdziarski} {et~al.}(2009){Zdziarski}, {Kawabata}, \&
  {Mineshige}}]{ZKM09}
{Zdziarski}, A.~A., {Kawabata}, R., \& {Mineshige}, S. 2009, \mnras, 399, 1633

\bibitem[{{Zdziarski} {et~al.}(2002){Zdziarski}, {Poutanen}, {Paciesas}, \&
  {Wen}}]{ZPP02}
{Zdziarski}, A.~A., {Poutanen}, J., {Paciesas}, W.~S., \& {Wen}, L. 2002, \apj,
  578, 357

\end{thebibliography}

\end{document}